# Unlocking the Potential: A Novel Tool for Assessing Untapped Micro-Pumped Hydro Energy Storage Systems in Michigan

Sharaf K. Magableh, Xuesong Wang, Oraib Dawaghreh, Caisheng Wang
*Department of Electrical and Computer Engineering, Wayne State University, Detroit, United States*

*Abstract*—This study presents an innovative tool designed to unlock the potential of Michigan's lakes and dams for applications such as water resource management and renewable energy generation. Given Michigan's relatively flat landscape, the focus is on systems that could serve as micro-hydro energy storage solutions. To ensure accuracy and reliability, the tool incorporates extensive data gathered from authorized sources, covering more than 420 water facilities and potential reservoirs in the state. These data are used as part of a case study to evaluate the tool's capabilities. Key parameters assessed include horizontal and vertical distances (head), volume, and the total storage capacity of each reservoir, measured in GWh. By analyzing these factors, the tool determines the suitability of various lakes and dams for hydroelectric power generation, and other uses based on the horizontal and vertical threshold distances. Its robust assessment framework integrates these metrics to comprehensively evaluate each site's potential. The tool's friendly interface and advanced data visualization features make the findings easy to interpret, facilitating optimal resource utilization and informed decision-making for state authorities. Hence, this tool represents a meaningful advancement in managing Michigan's water resources sustainably, promoting environmentally friendly practices, and supporting economic development.

NOMENCLATURE

*Abbreviations*
ESS     Energy Storage System
IEA     International Energy Agency
LR     Lower Reservoir
PSH     Pumped Storage Hydropower
RER     Renewable Energy Resource
UR     Upper Reservoir

*Symbols*
$E_{Total}$    Total Energy ($Joules$)
$E_{Storage}$   Potential energy Storage ($kWh$)
$V$     Volume of water reservoir ($m^3$)
$\rho$     Density of water, typically 1000 ($kg/m^3$)
$g$     Acceleration due to gravity, typically 9.81 ($m/s^2$)
$h$     Vertical elevation (head) between UR and LR ($m$)
$\overline{W}_{UR}$    Average width of the UR ($m$)
$\overline{H}_{UR}$    Average height of the UR ($m$)
$\overline{L}_{UR}$    Average length of the UR ($m$)

## I. INTRODUCTION

Renewable energy resources (RERs) will continue to play their role in energy generation for sustainability. However, their overall development and implementation also face inconsistent availability and intermittent operability as major barriers to their widespread applications. To overcome these challenges, energy storage systems (ESSs) have been introduced to balance the varying energy generation with demand [1]. ESSs store excess energy generated during periods of surplus and release it when demand exceeds generation, ensuring a swift response [2]. Roughly two-thirds of annual global power capacity additions come from solar and wind energy, while pumped storage hydropower (PSH) makes up around 96% of the global storage power capacity and 99% of the stored energy volume [3]. The rest of the electricity storage market is primarily occupied by batteries, such as home and electric vehicle batteries, which are more favorite compared to PSH for short-term storage needs, ranging from minutes to hours. However, PSH remains more cost-effective for large-scale energy storage projects lasting several hours to weeks [4]. Most existing PSH facilities are located along rivers and are used in conjunction with hydroelectric generation. During periods of low energy demand, water is pumped from a lower reservoir (LR) to an upper one, storing the energy for later use [5]. Interest is growing in closed-loop off-river PSH systems, where water is circulated between two closely spaced small reservoirs located away from natural waterways. These systems offer vast storage opportunities for the future [6]. Over the past ten years, literature reviews show that several studies on hybrid RERs have been conducted worldwide. This includes diverse assessments of ESSs like PSH. For instance, the authors in [1] reviewed several energy storage technologies (ESTs) and provided an overview of key performance metrics for major energy storage options. These metrics include efficiency, energy capacity and density, capital investment costs, lifespan, and self-discharge. Their findings suggest that selecting the appropriate storage system depends on specific needs, such as enhancing overall energy capacity and maintaining energy security. PSH has garnered special interest globally in recent years. The authors in [7] presented a comprehensive review of PSH systems, covering the basic principles, design considerations, and several configurations of PSH systems, including open and closed loops and hybrid designs. The paper provides insights into the challenges posed by PSH systems, highlighting their potential to contribute significantly to a sustainable and dependable energy future. To show the significance of ESS, the researchers in [8] proposed a grid-connected double storage system comprising PSH and batteries powered by solar PV and wind energy. In this hybrid setup, batteries capture surplus RERs that PSH cannot store and cater to loads beyond the water turbine's capacity. The system's energy management strategy accounts for situations where both PSH and batteries are charging by prioritizing battery storage, thus preventing renewable energy curtailment. If both storage systems reach their charging limits, excess energy can be fed

Sharaf K. Magableh, Xuesong Wang, Oraib Dawaghreh, and Caisheng Wang are with the Department of Electrical and Computer Engineering, Wayne State University, Detroit, USA (e-mail: sharaf.magableh@wayne.edu, xswang@wayne.edu, oraib.dawaghreh@wayne.edu, cwang@wayne.edu).

into the national grid. Findings reveal a 22.2% reduction in electricity costs by coupling batteries with PSH, with minimal reliance on the national grid for energy exchange.

The National Renewable Energy Laboratory (NREL) developed a tool for assessing PSH supply curves across the U.S [9]. This interactive map and geospatial dataset provide critical insights into PSH potential by quantifying the quantity, and cost of PSH resources at various sites. Users can explore sites with different storage durations, dam heights, and head heights, including repurposed open-pit mines and existing reservoirs, making this tool a robust resource for large-scale PSH site identification [10]. However, the NREL supply curves tool does not specifically address the potential of micro-PSH projects or smaller-scale applications in states like Michigan, where the geography and resource characteristics are better suited to micro-hydro systems. This limitation highlights the need for more localized tools, such as the micro-PSH assessment model developed in this study, to evaluate Michigan's unique water resources for smaller-scale, sustainable energy storage solutions. Hence, this study introduces a tool that assesses the viability of Michigan's lakes and dams for micro-PSH applications, focusing on renewable energy generation. Given Michigan's relatively flat geography, the research prioritizes smaller-scale hydro systems suited to the state's landscape, aiming to tap into currently underutilized water resources. In this paper, the tool evaluates over 420 potential sites by analyzing essential parameters such as horizontal and head, volume, and storage capacity (GWh). These insights will help identify reservoirs capable of supporting hydroelectric generation at a micro-scale, particularly for energy storage. While renewable energy research in Michigan often highlights wind and solar resources, micro-PSH studies are still new. Limited but notable work suggests that Michigan's abundant lakes and reservoirs could serve as ideal locations for micro-hydro or small-scale PSH installations, leveraging both natural elevation changes and infrastructure like dams [11]. Research on existing PSH systems underscores the importance of detailed geographic and hydro-resource assessments to optimize power output and sustainability [12]. However, Michigan-specific studies remain rare, signaling a research gap and an opportunity for innovation in this field.

## II. DATA COLLECTION, UTILIZATION AND MATHEMATICAL MODELING

This section outlines the data collection process, which is fundamental for the mathematical modeling used to estimate potential energy storage between reservoirs. By leveraging detailed lake metrics and horizontal distance measurements, the energy storage potential could be practically assessed.

### A. Data Integration

Michigan, known as the Great Lakes State, boasts an extraordinary abundance of water resources. It is home to four of the five Great Lakes, which collectively hold about 20% of the world's fresh surface water. Beyond these vast bodies, Michigan is dotted with over 11,000 inland lakes and numerous rivers and streams [13, 14]. This extensive network of water facilities not only provides recreational opportunities and supports biodiversity but also holds significant potential for energy storage solutions. The state's abundant reservoirs, combined with their varying elevations and volumes, present a promising opportunity for harnessing hydroelectric power and optimizing water management strategies [15]. This rich aquatic landscape makes Michigan an ideal case study for exploring and developing advanced energy storage systems based on water reservoir data.

Our dataset encompasses information for over 420 reservoirs in Michigan, that have a surface area of more than one $km^2$. Note that, the collected data includes critical metrics such as high-resolution surface elevation dataset above sea level and reservoir surface and bottom height information. These measurements allow us to calculate the depth of each reservoir accurately, which is essential for determining the volume of water available for potential energy storage. These data were obtained from officially authorized websites. This includes Shapefiles containing lake areas in square miles, geographic locations, and corresponding states [16], as well as surface elevation data [17]. and bottom elevation data [18]. This comprehensive data enables precise volume calculations, which are pivotal in assessing the energy storage capacity. Beyond individual lake metrics, our dataset also includes an extensive collection of horizontal distance measurements between these water facilities, totaling over 91,000 data points. This aspect of the data is crucial for our secondary objective of evaluating the feasibility of energy storage between different reservoirs. By analyzing the horizontal distances, we can identify pairs of reservoirs within practical proximity for energy transfer and storage. This extensive and detailed dataset forms the backbone of our analysis, enabling a robust assessment of potential energy storage solutions through both individual reservoir characteristics and inter-reservoir relationships.

### B. Mathematical Modeling

This paper employed straightforward mathematical modeling to estimate the potential energy storage between reservoirs, as illustrated in Fig. 1, focusing on a few key equations.

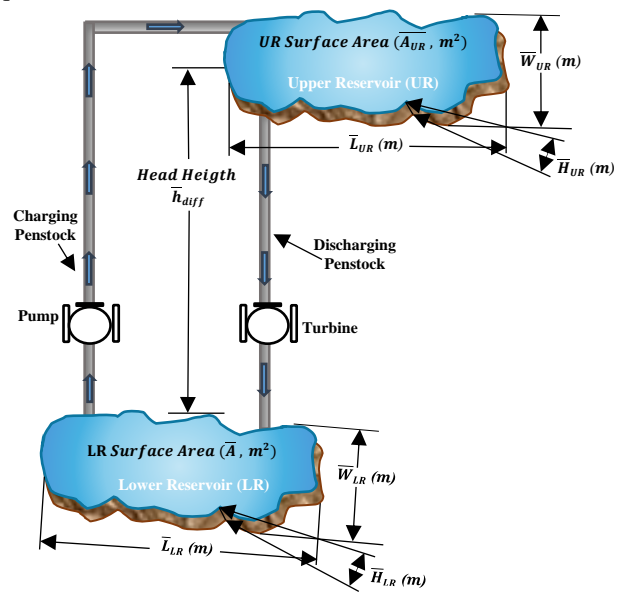

Fig. 1. Schematic diagram of the tool utilization for the PSH upper and lower reservoirs [19].

The simplicity of these equations underscores the fundamental relationship between water volume, gravitational force, and height difference in determining energy potential. Mathematical modeling and accurate data are crucial for this design methodology as they provide a systematic and quantifiable approach to evaluate and optimize energy storage solutions. Hence, the total energy $E_{Total}$ in Joules is calculated using (1). However, the energy storage in ($kWh$) is then determined by converting the total energy, as illustrated in (2).

$$E_{Total}(Joules) = V \times \rho \times g \times h \quad (1)$$

$$E_{Storage}(kWh) = \frac{E_{Total}}{3.6 \times 10^6} \quad (2)$$

Where; $V$ is the volume of the water reservoir as shown in Fig. 1, and can be computed as in (3). Accordingly, $\rho$ is the density of water $1000\ kg/m^3$, $g$ is the acceleration due to gravity; $9.81\ m/s^2$, and $h$ is the elevation of the head vertical difference between the upper reservoir (UR) and the LR. Note that, $\overline{A_{UR}}$ and $\overline{H}_{UR}$ represent the surface area and the average depth of the $UR$, respectively.

$$V = \overline{A_{UR}} \times \overline{H}_{UR} \quad (3)$$

## III. DESIGN METHODOLOGY

This project introduces a new MATLAB tool and graphical user interface (GUI) designed to analyze the potential of lakes and water facilities, considering Michigan as a case study. Indeed, implementing the collected data to estimate potential energy storage between reservoirs is crucial for future optimizing energy management and sustainability efforts. Hence, this project emphasizes developing and utilizing a MATLAB tool and GUI to improve accuracy and enhance user accessibility for these applications. This will also help researchers and engineers make initial decisions about the selected areas before proceeding with these real-life projects. The focus is on explaining the practical steps taken to gather and process relevant data for optimizing energy storage solutions. By gathering reliable data on water surface levels for more than 420 reservoirs, bottom levels, surface area, vertical distance, and horizontal distances between reservoirs, the tool standardizes units for consistency and performs necessary calculations. Users can input the name of a specific lake or reservoir, prompting the tool to search the dataset, extract relevant information, and convert measurements. The tool calculates average depth and volume, identifies distances under some threshold that defaults to $1km$, and determines potential energy storage. This is because there cannot be any effective reservoir if the horizontal distance between them is too far. However, this threshold could be changed based on the user demand. Results, including lake name, vertical distance, potential energy, and horizontal distance, are organized and stored in a cell array for easy access and analysis. Fig. 2 shows the flow chart for the proposed methodology implemented in this study.

Hence, the objective of this study is to develop a new tool and GUI using MATLAB to analyze the potential of lakes and water facilities. In addition to exploring the potential of unused water resources in the U.S., starting with Michigan as a case study. The flow chart of the method is shown in Fig. 2.

1) Data collection: Begin by gathering reservoir data from reliable sources. Filter the Michigan reservoirs to include only those with a surface area larger than $1km^2$, and perform a spatial join to calculate the minimum, maximum, and average surface and bottom elevations.
2) Assign unique reservoir IDs: Assign each reservoir a unique ID to enhance search accuracy, especially for those with identical names. These IDs are used to calculate the distances between reservoirs, resulting in over 91,000 distance measurements based on their coordinates.
3) Standardize data and define constants: Define the necessary constants and equations for the calculations.
4) Search functionality: Prompt the user to input a reservoir name. Utilize loops to search the dataset and extract key details such as surface area, elevation, and volume. Convert surface area from square miles to $km^2$ as needed.
5) Calculate depth, volume, and potential energy: Compute the average depth in m and the volume in $m^3$, then filter distances within the user-defined threshold. Calculate potential energy storage based on the vertical distance between reservoirs.
6) Organize and Display Results: Store and display the results, such as reservoir name, distances, and potential energy in a table for easy interpretation and further analysis.

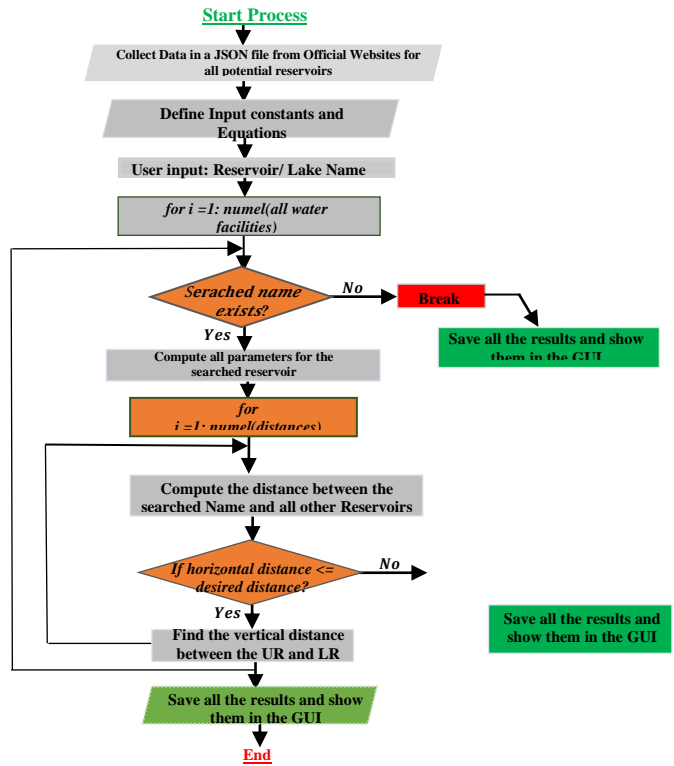

Fig. 2. Flow chart for the proposed methodology.

## IV. TOOL FEATURES, FUNCTIONALITY, AND CASE STUDY INSIGHTS

The "*Reservoir Info App*" shown in Fig. 3 is a user-friendly MATLAB application designed to help users explore and analyze reservoir data, with a particular focus on calculating potential energy storage between reservoirs. This includes more than 420 potential reservoirs with a surface area of more than one $km^2$, in Michigan as a case study.

This app features an intuitive interface user interface, enabling users to select reservoirs from a list easily or search by name if known, set threshold distances, and retrieve comprehensive information about each reservoir, such as surface area, average depth, volume, and potential energy storage capacity. Moreover, the application allows users to export the results to an Excel file for further analysis and documentation.

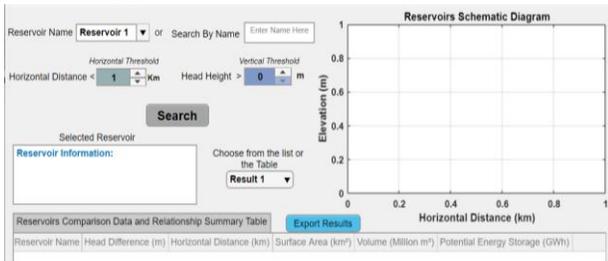

Fig. 3. The proposed UI tool for identifying reservoirs potential.

*A. Tool's Feature and Functionality*

Here's an overview of how the app behaves across different scenarios based on user inputs and interactions, highlighting its various features and functionalities.

- **Feature 1:** Intelligent Reservoir Search

This function performs a search for all data based on the user's query. It updates the table and information text areas with relevant data, based on the searched reservoir, including potential energy calculations between reservoirs within a specified threshold distance. If no exact match is found, the app suggests a potential alternative based on the closest matching name, using *'Levenshtein distance'* to measure similarity.

- **Feature 2:** Customizable Threshold Distances

Two spinners, one for the Horizontal Threshold and another for the Vertical Threshold, are introduced to set distance threshold values, as shown in Fig. 4.

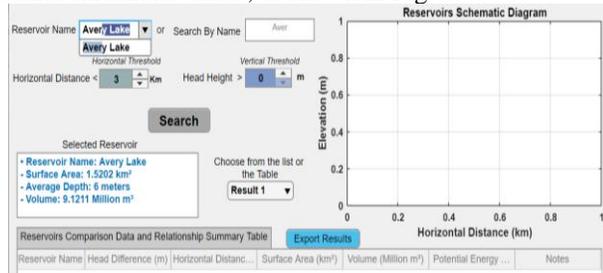

Fig. 4. Reservoir information app: key features and lake search by name.

The horizontal threshold spinner updates the threshold distance property when its value changes. For instance, when a user sets the threshold distance to 3*km*, the application then uses this value to filter and compute relevant data. Note that this threshold is initialized to be 1*km*, as it is recommended that both the UR and LR remain close to each other to efficiently pump or store water during charging or discharging modes [20].

- **Feature 3:** Interactive Table and Reservoir Visualization

The app allows users to interact with reservoir data through the UI Table Cell Selection feature. When a cell is selected, the app retrieves and displays key information, including the reservoir name, horizontal distance, surface area, volume, and vertical head difference between the UR and LR. Based on user-defined thresholds for horizontal distance and vertical elevation, the app determines which reservoir qualifies as the UR. It then computes the potential energy storage for the selected reservoirs.

For enhanced data visualization, the app generates a schematic plot that shows the relative distances and elevations of the reservoirs, as shown in Fig. 5. This visual representation helps users clearly understand the spatial relationship between reservoirs, facilitating decisions about optimal UR selection and energy storage calculations. Users can toggle between viewing potential energy storage directly in the table or accessing it from the drop-down result list. Additionally, specific notes for each row are displayed, and the table columns are automatically adjusted for better readability.

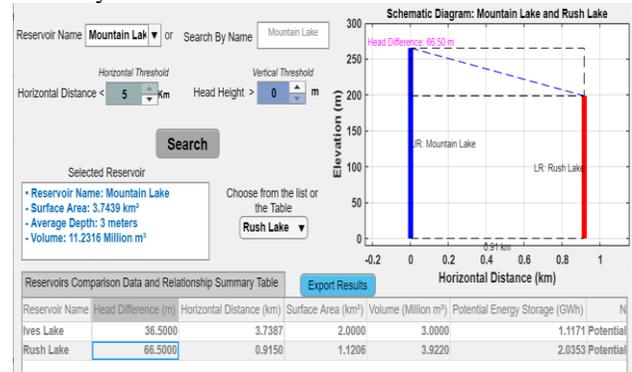

Fig. 5. Results of randomly searched reservoirs with the UI table and schematic diagram.

- **Feature 4:** Quick Export of Results

The app provides an "*Export Results*" function, allowing users to easily save reservoir data and calculations as an Excel file. After the export, a confirmation message is displayed, simplifying the process of saving and sharing data for further analysis.

*B. Reservoir Case Study and Analysis*

One of the key applications of PSH is in irrigation and maintaining a consistent supply of clean water. Indeed, by accurately determining potential energy storage, reservoir managers can more effectively plan the use of stored water for agriculture. Therefore, the motivation for this tool arises from the importance of identifying potential storage capacity for untapped PSH. To illustrate this tool's capabilities, let's use Lake Huron as an example. Known for its vast volume of about $1.393 \times 10^{12} \ m^3$, Lake Huron is the fifth-largest lake in the world in terms of surface area. When a user searches for Lake Huron, the tool will first display information about the lake. Then, based on the user-defined threshold distances, the tool will automatically generate a UI table, as in Fig. 6.

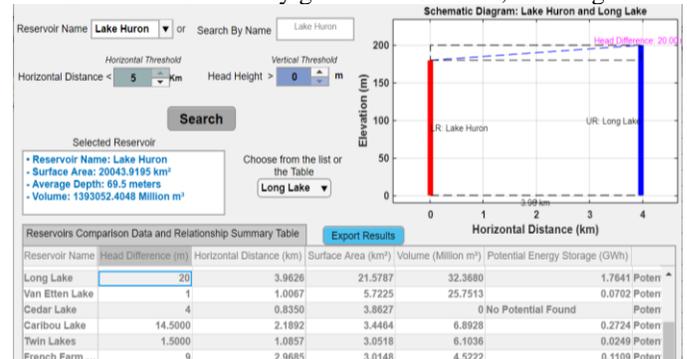

Fig. 6. Complete UI results of the tool for Lake Huron search.

The table lists all other potential lakes within the specified thresholds. The tool identifies nearby lakes based on their IDs, displaying data such as the head difference and horizontal distance between each reservoir. It also includes the surface area in $km^2$ and the volume in million $m^3$ for each listed reservoir, as illustrated in Fig. 7.

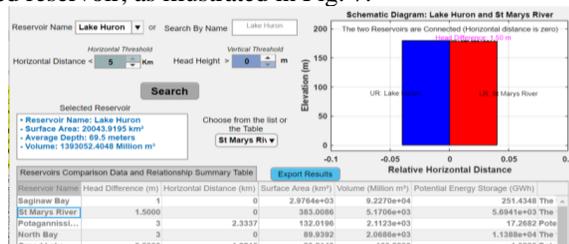

Fig. 7. Schematic diagram for the two connected reservoirs: Lake Huron and North Bay.

The tool then calculates which lake has the higher surface elevation above sea level to designate it as the potential UR. Subsequently, it computes the potential energy storage for the selected UR and generates a schematic diagram. Note that if the two reservoirs are connected, the relative horizontal distance will appear. Users can view the plots by selecting the corresponding row from the table or from the list. Finally, users can export and save the results to an Excel file for better readability and future reference.

## V. Conclusion

The "*Reservoir Info App*" in this study offers a powerful, data-driven tool to assess the untapped potential of Michigan's lakes and dams for micro-hydro energy storage and water resource management. By integrating comprehensive data sets and advanced computational methods, the app evaluates critical parameters like reservoir volume, surface distances, and potential energy storage in GWh, providing valuable insights for assessing hydroelectric power generation viability. This research could benefit projects that focus more on pre-evaluation, planning, and optimization studies, especially by exploring the potential of untapped hydropower reservoirs. The user-friendly interface and effective data visualization make it a valuable resource for a broad audience, including policymakers, researchers, and industry experts. This application allows users to find and evaluate nearby reservoirs, even identifying direct connections between them.

Future enhancements could amplify its impact by integrating real-time data and machine learning algorithms to predict reservoir behavior under various climatic conditions, increasing its robustness. Expanding coverage nationwide and keeping data updated would enhance its value. Additionally, incorporating advanced visualization tools like 3D modeling and virtual reality could enrich user interaction. Integrating the app with online platforms and other water resource management systems would foster collaborative strategies. Lastly, addressing high-resolution data needs and stakeholder feedback will support sustainable water management and renewable energy initiatives.

## Acknowledgment

This work was supported in part by the National Science Foundation of USA under Grant ECCS-2146615 and partially supported by the Department of Energy, Solar Energy Technologies Office (SETO) Renewables Advancing Community Energy Resilience (RACER) program under Award Number DE-EE0010413. Any opinions, findings, conclusions, or recommendations expressed in this material are those of the authors and do not necessarily reflect the views of the Department of Energy.